\documentclass[prd,aps,eqsecnum,twoside,twocolumn]{revtex4-1}

\usepackage{hyperref,amsfonts,amsmath,amssymb,graphicx,bm,subfigure}

\begin{document}

\author{Maxim Dvornikov}
\email{maxdvo@izmiran.ru}

\author{V.~B.~Semikoz}
\email{semikoz@yandex.ru}

\title{Evolution of axions in the presence of primordial magnetic fields}

\affiliation{Pushkov Institute of Terrestrial Magnetism, Ionosphere
and Radiowave Propagation (IZMIRAN),
108840 Moscow, Troitsk, Russia}
%
\begin{abstract}
We study the evolution of axions interacting with primordial magnetic fields (PMFs) starting just from the QCD phase transition in the expanding universe. This interaction is owing to the Primakoff effect. Adopting the zero mode approximation for axions, we derive the system of equations for axions and magnetic fields, where the expansion of the universe and the spectra of magnetic fields are accounted for exactly. We find that the contribution of the Primakoff effect to the dynamics of axions and magnetic fields is rather weak. It confirms some previous estimates leading to analogous conclusions, when accounting here for  the Hubble expansion both for an uniform axion field and non-uniform PMFs using Fourier spectra for their energy and helicity densities. We solve the corresponding system of the evolution equations and find that the axion zero mode, when evolving during radiation era, has its amplitude at the level sufficient for that axion to be a good candidate for the cold dark matter.
\end{abstract}
\maketitle

\section{Introduction}

As the solution to the CP problem in quantum chromodynamics (QCD), Peccei and Quinn suggested a mechanism that naturally gives rise to a Nambu-Goldstone boson, the so-called axion \cite{PQ,Weinberg:1977ma,Wilczek}. In addition, the axion also provides a
possible candidate for the cold dark matter (CDM) of the Universe~\cite{Rin12}. There are active searches for axions~\cite{Gra15}. A progress in the experimental studies of these particles is reported in Ref.~\cite{Apr20} recently.

The simultaneous presence of primordial magnetic fields (PMFs) and an axion field in a hot universe plasma near the QCD phase transition (QCDPT) allows to study their mutual influence during the radiation era. We adopt that in the universe cooling $T_\mathrm{PQ}\to T_\mathrm{QCD}$, where $T_\mathrm{PQ}\simeq f_a=(10^{10}\div 10^{12})\,{\rm GeV}$ is the temperature at the Peccei-Quinn phase transition, $T_\mathrm{QCD}\simeq (100\div 200)\,{\rm MeV}$ is the temperature at QCDPT, and $f_a$ is the Peccei-Quinn parameter, a small axion mass $m_a(T)$ increases and gets after mixing with $\pi^0$ mesons its ``cold'' (fixed) value $m_a(0)=m_a$ just near the initial time  in our problem $t_0=t_\mathrm{QCD}$,
\begin{equation}\label{mass}
m_a=6\times 10^{-6}\,{\rm eV}\left(\frac{10^{12}\,{\rm GeV}}{f_a}\right).
\end{equation}
In its turn, PMFs  originate from the hypermagnetic fields $B_\mathrm{Y}$ which exist before the electroweak phase transition (EWPT), $T> T_\mathrm{EW}\simeq (100\div 160)\,{\rm GeV}$. Such PMF, being a seed for observable galactic magnetic fields $B\sim 10^{-6}\,{\rm G}$, can be strong enough at $T_\mathrm{QCD}$. Returning the present bound on the extragalactic PMF, $B\leq B_\mathrm{now}=10^{-9}\,{\rm G}$, back to early universe times, one gets the PMF strength $B_\mathrm{QCD}=B_\mathrm{now}(1 + z_\mathrm{QCD})^2\simeq 1.6\times 10^{14}\,{\rm G}$ for the red shift $z_\mathrm{QCD}=4\times 10^{11}$ at $T_\mathrm{QCD}\sim 100\,{\rm MeV}$, or $B_\mathrm{EW}=1.6\times 10^{20}\,{\rm G}$ at $T_\mathrm{EW}\sim 100\,{\rm GeV}$. 

On one hand, such a strong field  can not influence the Universe expansion since the PMF energy density $\rho_\mathrm{B}=B^2/2$ is much less than the ultrarelativistic matter energy density $\rho=g^*\pi^2T^4/30$, $\rho_\mathrm{B}\ll \rho$. Here $g^*$ is the effective number of degrees of freedom~\cite[p.~95]{Gorbunov:2011zz}. On the other hand, it is intriguing to check whether such a field can be enlarged by the axion instability term $\sim \dot{\varphi}\nabla\times {\bf B}$ entering the induction equation \cite{Long:2015cza},
\begin{equation}\label{Faraday}
\frac{\partial {\bf B}}{\partial t}=\frac{\nabla^2{\bf B}}{\sigma_\mathrm{cond}} + \frac{g_{a\gamma}}{\sigma_\mathrm{cond}}\frac{{\rm d}\varphi}{{\rm d}t}\nabla\times {\bf B},
\end{equation}
or, vice versa, the PMF can influence evolution of that uniform axion field, $\nabla\varphi=0$,
\begin{equation}\label{zero_mode}
\ddot{\varphi} + m_a^2\varphi= \frac{g_{a\gamma}}{\sigma_\mathrm{cond}}(\nabla\times {\bf B})\cdot{\bf B}.
\end{equation}
Here $g_{a\gamma}\approx\alpha_\mathrm{em}/2\pi f_a$ is the coupling constant for axion interaction with electromagnetic fields, $\alpha_\mathrm{em}\approx(137)^{-1}$ is the fine structure constant, $\sigma_\mathrm{cond}\simeq 100 T$ is the electric conductivity in the hot universe plasma.

The problem of the mutual influence of axions and PMFs was discussed in Ref.~\cite{Long:2015cza}. However, in Ref.~\cite{Long:2015cza}, the Hubble expansion was neglected for both a uniform axion gas and primordial magnetic fields. We study the similar problem accounting for both that expansion when using conformal variables and an inhomogeneity of PMFs through their Kolmogorov spectra for the densities of the magnetic energy and the magnetic helicity.

Our work is organized as follows. In Sec.~\ref{sec:AXEARLYUNIV}, we present the complete system of the evolution equations for the homogeneous axion field $\varphi (\eta)$ and the spectra of the conformal PMF energy density $\rho_c(k_c,\eta)$ and the conformal PMF helicity density $h_c(k_c,\eta)$. Here $\eta$ is the conformal time and $k_c$ is the conformal momentum. Details for derivation of such a system are given in Appendix~\ref{A}. In Sec.~\ref{sec:INICOND}, we formulate the initial conditions for this system of evolution equations. In Sec.~\ref{sec:RES}, we present the results of the numerical calculations
illustrated by plots both for the maximum extragalactic field $B_\mathrm{now}=10^{-9}\,{\rm G}$ and without PMFs at all. In Sec.~\ref{sec:DISC}, we discuss our results.

\section{Axions in the early universe plasma with PMFs\label{sec:AXEARLYUNIV}}

For a homogeneous axion field   $\nabla \varphi=0$, or for the zero axion mode~\cite{f1}, the axion evolution equation in the presence of PMF and in the Friedmann-Robertson-Walker (FRW) metric, $\mathrm{d}s^2=a^2(t)(\mathrm{d}\eta^2 - \mathrm{d}{\bf x}^2)$, where $a$ is the scale factor and $\mathrm{d}\eta=dt/a(t)$,  takes the following form in the variables $x^{\mu}=(\eta, {\bf x})$ (see Eqs.~(\ref{eq:phic}) and~(\ref{eq:hc}) in Appendix~\ref{A}):
\begin{equation}\label{axion_wave}
\frac{\mathrm{d}^{2}\varphi}{\mathrm{d}\eta^{2}}+\frac{2}{a}\frac{\mathrm{d}a}{\mathrm{d}\eta}\frac{\mathrm{d}\varphi}{\mathrm{d}\eta}+m_a^{2}a^{2}\varphi=-\frac{g_{a\gamma}}{2a^2}\frac{\mathrm{d}h_c(\eta)}{\mathrm{d}\eta}.
\end{equation}
Here $h_c(\eta)=a^3h(t)=V^{-1}\int \mathrm{d}^3x{\bf A}_c(\eta,{\bf x})\cdot{\bf B}_c(\eta,{\bf x})$ is the conformal PMF helicity density, ${\bf A}_c(\eta,{\bf x})=a{\bf A}$ is the vector potential,~${\bf B}_c(\eta,{\bf x})=\nabla\times{\bf A}_c(\eta,{\bf x}) =a^2{\bf B}$ is the comoving magnetic field in conformal variables~\cite{f2}. 

We can calculate the derivative $\mathrm{d}h_c(\eta)/\mathrm{d}\eta$ for the magnetic helicity density $h_c(\eta)=\int \mathrm{d}k_c h(\eta, k_c)$ given by its isotropic Fourier spectrum $h_c(k, \eta )=k^2_c{\bf A}_c(k_c,\eta){\bf B}_c^*(k_c,\eta)/(2\pi^2V)$ as $\partial_{\eta}h_c(\eta)=\int \mathrm{d}k_c\partial_{\eta}h_c(\eta,k_c)$, where $k_c=ak=\text{const}$ is the comoving momentum that does not depend on time. We use the evolution equations for PMF spectra for this purpose (see Appendix~\ref{A}),
\begin{align}
\frac{\partial h_{c}(k_{c},\eta)}{\partial\eta} & =-\frac{2k_{c}^{2}}{\sigma_{c}}h_{c}(k_{c},\eta)+\frac{4g_{a\gamma}}{\sigma_{c}}\frac{\mathrm{d}\varphi(\eta)}{\mathrm{d}\eta}\rho_{c}(k_{c},\eta),
  \label{spectrah}
  \\
\frac{\partial\rho_{c}(k_{c},\eta)}{\partial\eta} & =-\frac{2k_{c}^{2}}{\sigma_{c}}\rho_{c}(k_{c},\eta)+k_{c}^{2}\frac{g_{a\gamma}}{\sigma_{c}}\frac{\mathrm{d}\varphi(\eta)}{\mathrm{d}\eta}h_{c}(k_{c},\eta),
  \label{spectra}
\end{align} 
where the magnetic energy density $\rho_c(\eta)=\int \mathrm{d}k_c \rho_c (\eta, k_c)=B^2_c(\eta)/2$ is given by the isotropic Fourier spectrum $\rho_c(k_c, \eta)=k_c^2{\bf B}_c(k_c,\eta){\bf B}_c^*(k_c,\eta)/(4\pi^2V)$, $k_c=|{\bf k}_c|$ is the absolute value of the comoving Fourier momentum, and $\sigma_c=a\sigma_\mathrm{cond}$ is given by the electric conductivity $\sigma_\mathrm{cond}\simeq 100 T$ in the hot universe plasma.

Substituting the derivative $\mathrm{d}h_c(\eta)/\mathrm{d}\eta=\int_{k_\mathrm{min}}^{k_\mathrm{max}} \mathrm{d}k_c\partial_{\eta}h_c(\eta,k_c)$, with the integrand given by Eq.~(\ref{spectrah}), we can rewrite the axion evolution Eq.~(\ref{axion_wave}) as
\begin{align}
\frac{\mathrm{d}^{2}\varphi(\eta)}{\mathrm{d}\eta^{2}}+ & \frac{2}{a(\eta)}\frac{\mathrm{d}\varphi(\eta)}{\mathrm{d}\eta}\left(\frac{\mathrm{d}a(\eta)}{\mathrm{d}\eta}+\frac{g_{a\gamma}^{2}}{a(\eta)\sigma_{c}}\int_{k_{\mathrm{min}}}^{k_{\mathrm{max}}}\rho_{c}(k_{c},\eta)\mathrm{d}k_{c}\right)\nonumber \\
 & +m_a^{2}a^{2}(\eta)\varphi(\eta)=\frac{g_{a\gamma}}{a^{2}(\eta)\sigma_{c}}\int_{k_{\mathrm{min}}}^{k_{\mathrm{max}}}k_{c}^{2}h_{c}(k_{c},\eta)\mathrm{d}k_{c},\label{axion_wave2}
\end{align}
that is the generalization of the axion wave Eq.~(40a) in Ref.~\cite{Long:2015cza} accounting for the Hubble expansion and the Fourier spectra of non-uniform electromagnetic fields. 

The system of equations that describes an evolution of the axion field in PMF, given by Eq.~(\ref{axion_wave2}), is completed by the differential Eqs.~\eqref{spectrah} and~(\ref{spectra}) for the spectra $h_c(k_c,\eta)$ and $\rho_c(k_c,\eta)$. Here, in the causal scenario, the momentum limits $k_\mathrm{min}=k_1a(t)$ and $k_\mathrm{max}=k_2a(t)$ obey the inequality $k_2> k_1=l_\mathrm{H}^{-1}(T_0)=\text{const}$ where the horizon size $l_\mathrm{H}(T_0)$ is chosen at the QCD phase transition time $t_0= (T_0/{\rm MeV})^{-2}\,\mathrm{s}$ when the axion mass $m_a(T)$ reaches its maximum (``cold'') value, given in Eq.~(\ref{mass}), during cooling of the Universe, $T\to T_0$.
We consider such a time $t_0\simeq 10^{-4}\,\mathrm{s}$ as the initial moment in our simulations. Correspondingly, for the conformal time during radiation epoch, $\eta (t)=\int_{t_0}^{t}\mathrm{d}t/a(t) + \eta (t_0)$, for which $a(t)=a(t_0)\sqrt{t/t_0}$ is the scale factor in FRW metric, we put the initial value $\eta (t_0)=0$ resulting in
\begin{equation}\label{conformal}
\eta(t)=\frac{2t_0}{a(t_0)}\left[\sqrt{\frac{t}{t_0}} -1\right],
\end{equation}
where the initial scale factor $a(t_0)=T_\mathrm{now}/T_0=(1 + z_\mathrm{QCD})^{-1}$ is given by by the red shift at the QCD phase transition, $z_\mathrm{QCD}\simeq 4\times 10^{11}$, since $T_\mathrm{now}=2.725\,\mathrm{K}$ and $T_0=T_\mathrm{QCD}= 10^{12}\,\mathrm{K}\simeq 100\,{\rm MeV}$. 

We assume that PMF survives against the ohmic diffusion in the early Universe in a wide region of Fourier momenta $k_1<k_2$. Obviously, such a field survives for the minimal and fixed value $k_1=l_\mathrm{H}^{-1}=(2/3)\times 10^{-20}\,{\rm GeV}$ given by the horizon size $l_\mathrm{H}$, or maximum PMF scale $\Lambda^{(\mathrm{max})}_\mathrm{B}=l_\mathrm{H}$ at the initial time $t_0$. Additionally, we demand that the expansion time $t=t_0(T_0/T)^2$ be faster than the diffusion time $t_\mathrm{diff}=\sigma_\mathrm{cond}(\Lambda^{(\mathrm{min})}_\mathrm{B})^2$, or $t< t_\mathrm{diff}$ even for the minimal PMF scale $\Lambda^{(\mathrm{min})}_\mathrm{B}=k_2^{-1}$. For the initial time $t_0$ this means
\begin{equation}\label{kmax}
t_0\leq \frac{\sigma_c}{a(t_0)k_2^2},~~~~T\leq T_0,
\end{equation}
where the conductivity $\sigma_\mathrm{cond}=\sigma_c/a(t_0)$ was substituted.

The system of the evolution Eqs.~(\ref{spectra})-(\ref{axion_wave2}), rewritten for the three dimensionless functions $\mathcal{H}(\kappa, \tau)=g_{a\gamma}^2h_c(k_c,\eta)/2$, $\mathcal{R}(\kappa,\tau)=(g_{a\gamma}^2/k_\mathrm{max})\rho_c(k_c,\eta)$, $\Phi( \tau)=2(k_\mathrm{max}g_{a\gamma}/\sigma_c)\varphi(\eta)$ with change of the conformal time $\eta $ to the dimensionless one, $\tau=2k_\mathrm{max}^2\eta/\sigma_c$, reads,
\begin{align}
\label{systemH}
\frac{{\rm d}\mathcal{H}}{{\rm d}\tau}= & -\kappa^2\mathcal{H}(\kappa,\tau) + \frac{{\rm d}\Phi(\tau)}{{\rm d}\tau}\mathcal{R}(\kappa,\tau)
\\
\label{systemR}
\frac{{\rm d}\mathcal{R}}{{\rm d}\tau}= & -\kappa^2\mathcal{R}(\kappa,\tau) + \kappa^2\frac{{\rm d}\Phi (\tau)}{{\rm d}\tau}\mathcal{H}(\kappa,\tau)
\\
\notag
\frac{{\rm d}^2\Phi}{{\rm d}\tau^2} & + \frac{2}{a(\tau)}\frac{{\rm d}\Phi}{{\rm d}\tau}\left(\frac{{\rm d}a(\tau)}{{\rm d}\tau} + \frac{1}{2a(\tau)}\int_{\kappa_m}^1\mathcal{R}(\kappa,\tau)d\kappa\right)
\\
\label{systemPhi}
&  + a^2\mu^2\Phi (\tau) = \frac{1}{a^2(\tau)}\int_{\kappa_m}^1\kappa^2\mathcal{H}(\kappa,\tau)d\kappa .
\end{align}
Here, in Eq.~\eqref{systemPhi}, $\mu=m_a\sigma_c/2k_\mathrm{max}^2$ is the dimensionless parameter given by the axion mass in Eq.~(\ref{mass}). The lower limit $\kappa_m=k_\mathrm{min}/k_\mathrm{max}=k_1/k_2$ for the dimensionless Fourier momentum $\kappa=k_c/k_\mathrm{max}=k/k_2$ obeys the \text{const}raints
\begin{equation}\label{width}
  2.6\times 10^{-11}\leq \kappa_m\ll 1,
\end{equation}
where we account for Eq.~({\ref{kmax}).

Note that we have three free parameters for PMF in our problem:
\begin{itemize}
  \item a wide width of the Fourier spectrum for PMF given by a small varying  
  parameter $\kappa_m$ in Eq.~(\ref{width}), or by varying
  $k_\mathrm{max}=k_\mathrm{min}/\kappa_m$ for the fixed
  $k_\mathrm{min}=a(t_0)k_1$;
  \item the helicity parameter $q$, $0\leq | q|\leq 1$,
  entering the initial magnetic helicity spectrum
  (see Eq. (\ref{initial_spectrum}) below), where $q=0$ corresponds
  to the non-helical initial PMF, whereas $| q|=1$ to the maximum helical PMF; 
  \item an undetermined cosmological magnetic field $B_\mathrm{now}$
  at the present red shift $z=0$ originated in our scenario from the initial
  PMF $B_0=B_\mathrm{now}(1 +z_\mathrm{QCD})^2$ at
  $z_\mathrm{QCD}=4\times 10^{11}$ in the early Universe.
\end{itemize}
These three parameters fully describe necessary characteristics of PMF: the spatial scale, its structure (the topology) and the strength. 

Using the definition of the conformal time in Eq.~(\ref{conformal}), the scale in the radiation era, $a(t)=a(t_0)\sqrt{t/t_0}=a(t_0)[1 + a(t_0)\eta(t)/2t_0]$, can be rewritten in the new variables as $a(\tau)=a(t_0)[1 + K\tau]\equiv a(t_0)\tilde{a}(\tau)$, with the factor
\begin{equation}
  \tilde{a}(\tau)=1 + K\tau,
  \quad
  K=3.7\times 10^{20}\kappa_m^2,
\end{equation}
where $\tau\geq 0$~\cite{f3}.

\subsection{Axion energy density}

The important characteristic of axions is the axion energy density $\rho_a(t)=T_{00}$, where $T_{00}$ is the component of the energy-momentum tensor for a scalar field,
\begin{equation}\label{eq:Tmunu}
  T_{\mu\nu}=\partial_{\mu}\varphi\partial_{\nu}\varphi-
  \frac{g_{\mu\nu}}{2}
  \left(
    g^{\lambda\rho}\partial_{\lambda}\varphi\partial_{\rho}\varphi-
    m_a^{2}\varphi^{2}
  \right),
\end{equation}
In case of a spatially homogeneous field, basing on Eq.~\eqref{eq:Tmunu}, one gets that
\begin{align}\label{energy}
  \rho_a (t) =&\frac{1}{2} [\dot{\varphi}^2 + m_a^2\varphi^2]=\frac{1}{2}  
  \left(\frac{k_\mathrm{max}}{g_{a\gamma}a}\right)^2
  \notag
  \\
  & \times
  \left[\left(\frac{{\rm d}\Phi}{{\rm d}\tau}\right)^2 +
  \mu^2a^2\Phi^2(\tau)\right],
\end{align}
where $k_\mathrm{max}=k_2a=k_1a/\kappa_m$ is given by the parameter $\kappa_m$ in Eq.~(\ref{width}) and the minimal momentum $k_1$ in our causal scenario at $t_0=t_\mathrm{QCD}$, $k_1=l_\mathrm{H}(t_0)^{-1}=(2/3)\times 10^{-20}\,\text{GeV}$. This results in the dimensional factor ahead the brackets in Eq. (\ref{energy}), 
\begin{equation}
\label{constant}
C_m=\frac{1}{2}\left(\frac{k_\mathrm{max}}{g_{a\gamma}a}\right)^2= \frac{2}{9} \left(\frac{10^{-10}}{\kappa_m^2}\right)\,\text{GeV}^4,
\end{equation}
that depends on the PMF spectrum width $\kappa_m$.

\subsection{Initial conditions\label{sec:INICOND}}

The initial axion zero mode amplitude  $\varphi (t_0)\simeq f_a$, $\bar{\theta}(t_0)\simeq 1$, see, e.g., Ref.~\cite{Long:2015cza}, corresponds to 
\begin{equation}\label{initial_Phi}\Phi(\tau=0)=\frac{2k_\mathrm{max}g_{a\gamma}}{\sigma_c}\varphi (t_0)=\frac{1.5\times 10^{-24}}{\kappa_m},
\end{equation}
where $\kappa_m$ is the free parameter \text{const}rained in Eq.~(\ref{width}). 

The initial derivative $\mathrm{d}\Phi/\mathrm{d}\tau|_{\tau=0}$, necessary for solving Eq.~(\ref{systemPhi}), is derived using the virial theorem $\langle\dot{\theta}^2\rangle=m_a^2\langle \theta^2\rangle$ resulting from the axion energy density in Eq. (\ref{energy}) around its potential minimum \cite{Duffy:2009ig},
\begin{equation}
  \rho=\frac{f_a^2}{2}[\dot{\theta}^2 + m_a^2(t)\theta^2],
  \quad
  \theta (t)=\frac{\varphi}{f_a}.
\end{equation}
We put roughly $({\rm d\varphi}/{\rm d}t)_{t=t_0}=m_a\varphi (t_0)$ relying on that virial theorem. Then, one gets
\begin{equation}\label{derivative_axion}
  \left.
    \frac{{\rm d}\Phi}{{\rm d}\tau}
  \right|_{\tau=0}=\left(\frac{g_{a\gamma}a(t_0)}{k_\mathrm{max}}\right)m_af_a=10^3\kappa_m,
\end{equation}
where we substituted the ``cold'' axion mass $m_a$ in Eq.~(\ref{mass}) at the moment of the QCD phase transition, $m_a(T)\to m_a$ at $T\to T_\mathrm{QCD}$.

Note that the mass term, $(\mu a^2)\Phi$, in Eq.~(\ref{systemPhi})
\begin{equation}\label{mass2}
  (\mu a)^2=2.85\times 10^{30}\kappa_m^2(1 + K\tau)^2.
\end{equation}
is rather great. We shall see in Sec.~\ref{sec:RES} that this term is dominant.

The initial PMF spectra $\mathcal{R}(\kappa,\tau=0)$ and $\mathcal{H}(\kappa,\tau=0)$,
\begin{eqnarray}\label{spectra_cal}
&&\mathcal{R}(\kappa,\tau=0)=\frac{g_{a\gamma}^2}{k_\mathrm{max}}\rho_c(k_c,\eta=0),\\&&
\mathcal{H}(\kappa,\tau=0)=\frac{g_{a\gamma}^2}{2}h_c(k_c,\eta=0),\label{spectra_cal1}
\end{eqnarray}
are given by the PMF spectra in the known form~\cite{f4},
\begin{align}\label{initial_spectrum1}
  \rho_c(k_c,\eta=0) = & Ck_c^{n_\mathrm{B}},
  \notag
  \\
  C= & \frac{(n_\mathrm{B} + 1)B_\mathrm{now}^2}
  {2k_\mathrm{max}^{n_\mathrm{B} +1}(1 - \kappa_m^{n_\mathrm{B} + 1})},
  \\
  \label{initial_spectrum}
  h_c(k_c,\eta=0) = & \frac{2q}{k_c}\rho_c(k_c, \eta=0),
  \quad
  0\leq | q| \leq 1. 
\end{align}
The constant $C$ for the spectrum in Eq.~(\ref{initial_spectrum1}) is given by normalization on the initial magnetic energy density $B_c^2(\eta=0)/2=\int \mathrm{d}k_c\rho_c(k_c,\eta=0)$ for the comoving PMF strength $B_c(\eta=0)=B_\mathrm{now}$.
Thus, substituting Eqs.~(\ref{initial_spectrum1}) and~(\ref{initial_spectrum}) into Eqs.~(\ref{spectra_cal}) and~(\ref{spectra_cal1})
correspondingly we obtain the initial PMF spectra
\begin{align}\label{initial_cal}
\mathcal{R}(\kappa,\tau=0)=& \frac{C_0(n_\mathrm{B} +1)}{2(1 - \kappa_m^{n_\mathrm{B}+1})}\kappa^{n_\mathrm{B}},
\\
\mathcal{H}(\kappa,\tau=0)= & q\frac{C_0(n_\mathrm{B} +1)}{2(1 - \kappa_m^{n_\mathrm{B}+1})}\kappa^{n_\mathrm{B}-1},
\notag
\\
C_0=&\left(\frac{g_{a\gamma}B_\mathrm{now}}{k_\mathrm{max}}\right)^2.
\end{align}
We consider below the Kolmogorov's spectrum with $n_\mathrm{B}= - 5/3$.

\section{Results\label{sec:RES}}

In this section, we present the numerical solution of Eqs.~\eqref{systemH}-\eqref{systemPhi} with the initial conditions fixed in Sec.~\ref{sec:INICOND}.

We rely below in Fig.~\ref{fig:axion1},  on a maximum value for PMF at present time $B_\mathrm{now}=10^{-9}\,{\rm G}$ that obeys the bounds
\begin{itemize}
  \item $B_{1\,\mathrm{Mpc}}< 4.4\times 10^{-9}\,{\rm G}$ found in Ref.~\cite{Plank} 
  from CMB anisotropies (effect on CMB polarization induced by Faraday 
  rotation)  for zero magnetic helicity ($q=0$);
  \item $B_{1\,\mathrm{Mpc}}< 5.6\times 10^{-9}\,{\rm G}$ for a maximally helical   
  magnetic field ($q=1$).
\end{itemize}

Notice that, at the initial time $t_0=t_\mathrm{QCD}$ chosen in our problem for the causal scenario, the maximum spatial PMF scale $\Lambda_\mathrm{B}^{(\mathrm{max})}=l_\mathrm{H}(t_0)\simeq 3\times 10^6\,\mathrm{cm}$ occurs too small after its following growth till present time. It grows upto the scale $\Lambda_\mathrm{B}(z=0)\simeq 1\,\mathrm{pc}$ only since the horizon expands during the cooling much faster, $l_\mathrm{H}\sim T^{-2}$, than the correlation length $\Lambda_\mathrm{B}\sim T^{-1}$. However, the inverse cascade in the relativistic MHD, accounting for non-linear terms in Navier-Stokes equation (we do not touch here the full MHD approach), can rearrange the Fourier MHD spectra in such a way that $\Lambda_\mathrm{B}$ as a measure of the coherence length of the magnetic field, can increase in the coherence by the five orders of magnitude~\cite{Brandenburg:1996fc}. This fact allows us to get closer to the present scale $\Lambda_\mathrm{B}\sim 1\,{\rm Mpc}$ for PMF bounded in Ref.~\cite{Plank}.  The other problem is that, around the time of the recombination, the
photon diffusion becomes very large and so-called the Silk mechanism could destroy the PMF characteristics.  This danger was refuted in Ref.~\cite{Brandenburg:1996sa}, where nonlinear effects were shown to prevent most likely this problem from happening.

In Fig.~\ref{fig:axion1}, the normalized axion field $\Phi(\tau)/\Phi (0)=\varphi (t)/\varphi (t_0)=\bar{\theta}(t)$, starting from $\bar{\theta}(t_0)=\varphi (t_0)/f_a=1$, grows forty times just at the beginning (see inset in Fig. \ref{fig:axion1}), and then reduces soon to the same value $\bar{\theta} (t)\sim 1$ acceptable for a future axion contribution to CDM. This result does not depend on the presence of PMF, see in Fig.\ref{fig:axion4} where we put $g_{a\gamma}=0$ and $q=0$ excluding interaction of axions with the PMF. In both cases, this evolution of $\bar{\theta}(t)$-amplitudes happens during lepton era in the hot universe plasma.

\begin{figure*}
  \centering
  \subfigure[]
  {\label{fig:axion1}
  \includegraphics[scale=.38]{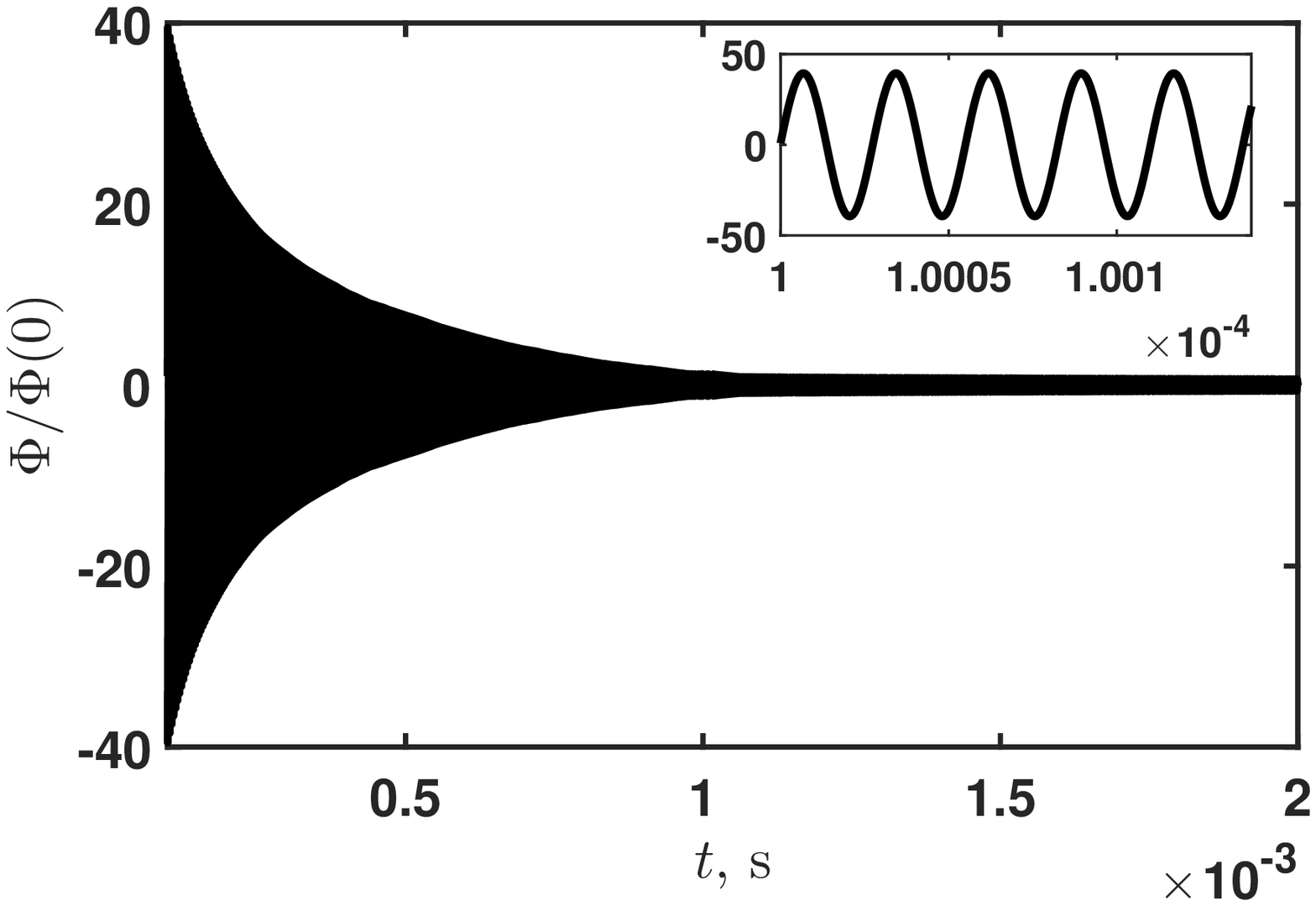}}
  \hskip-.6cm
  \subfigure[]
  {\label{fig:axion4}
  \includegraphics[scale=.38]{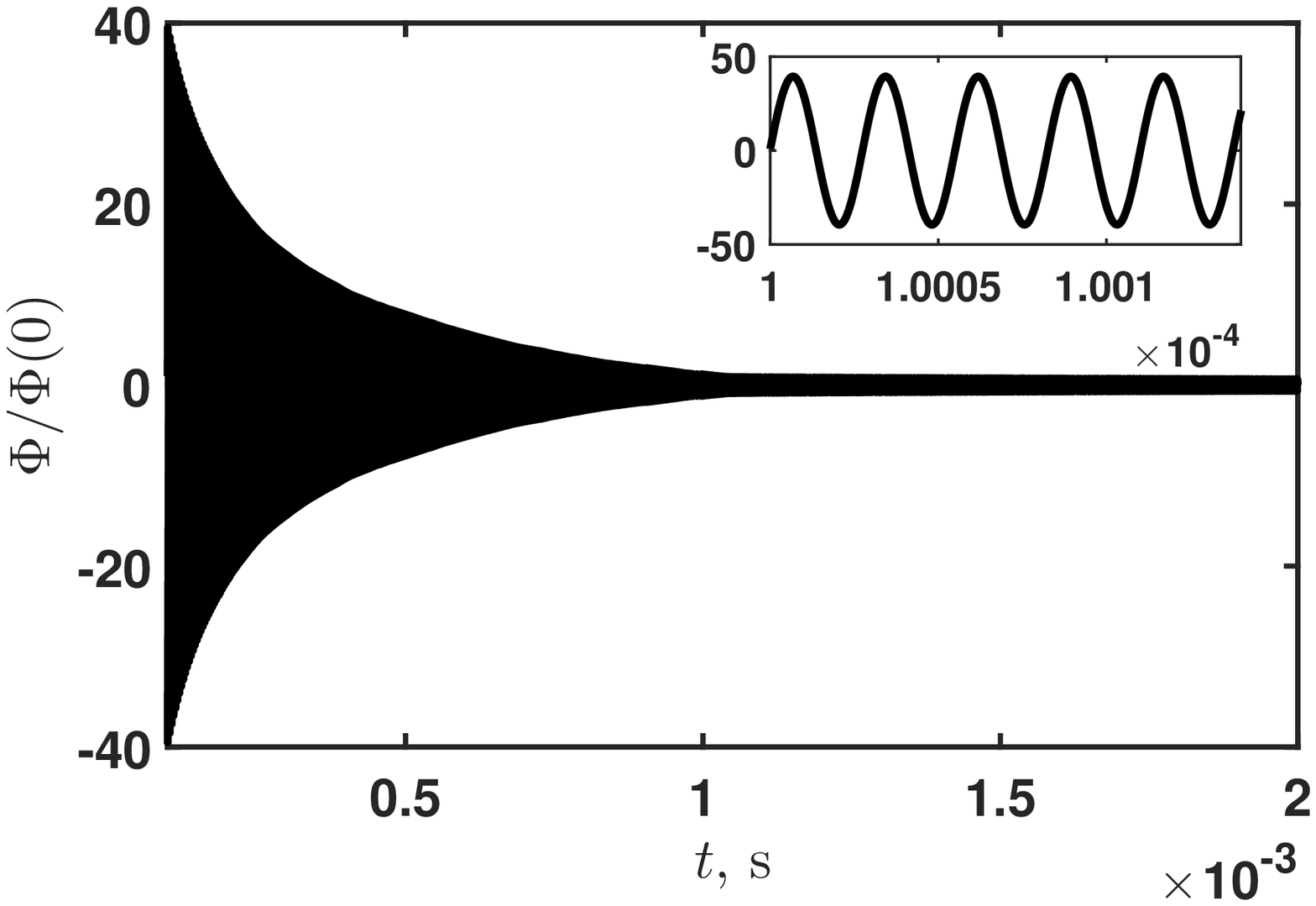}}
  \protect
  \caption{(a) The evolution of the axion zero mode amplitude $\Phi(\tau)/\Phi(\tau=0)\equiv\varphi (t)/f_a=\bar{\theta}(t)$ for the Peccei-Quinn parameter $f_\mathrm{PQ}\simeq f_a=10^{12}\,{\rm GeV}$ and the Fourier spectrum width  $\kappa_m= 10^{-10}$ in Eq.~(\ref{width}), choosing the maximal $B_\mathrm{now}=10^{-9}\,{\rm G}$. The seed magnetic field is supposed to be nonhelical, i.e. $q=0$ in Eq.~(\ref{initial_spectrum}). (b) The evolution of the axion zero mode amplitude $\Phi(\tau)/\Phi(\tau=0)\equiv\varphi (t)/f_a=\bar{\theta}(t)$ for the Peccei-Quinn parameter $f_\mathrm{PQ}\simeq f_a=10^{12}\,{\rm GeV}$ without PMF contribution, $g_{a\gamma}=0$ and $q=0$.\label{fig:axion}}
\end{figure*}

One can explain the appearance of a great amplitude $\Phi (\tau)/\Phi(0)\sim 40$ just after the start at $t\geq t_0=10^{-4}\,\mathrm{s}$ using initial conditions in Eqs. (\ref{initial_Phi})-(\ref{mass2}). For a slow change of the scale factor $a(\tau)$ and neglecting the PMF terms, we get oscillations of the axion field,  $\Phi (\tau) = \mathcal{A}\cos\omega\tau + \mathcal{B}\sin \omega\tau$ obeying the simplified equation in Eq.~(\ref{systemPhi}), $\Phi_{\tau\tau} + \omega^2\Phi(\tau)=0$,  $\omega =\mu a $, $\tau\ll 1$. Since $\mathcal{A}=\Phi (0)$ and $\mathcal{B}= \mathrm{d}\Phi/\mathrm{d}\tau|_{\tau=0}/\omega$, one finds from the initial conditions the corresponding ratio $\Phi(\tau)/\Phi (0)$ for the amplitude of $\Phi (\tau)$ at $\omega\tau=\pi/2$,
\begin{align}
\frac{\Phi (\tau)}{\Phi (0)}=& \frac{\mathcal{B}}{\mathcal{A}},
\quad
\text{or}
\quad
\frac{\Phi (\tau)}{\Phi (0)}=\frac{(\mathrm{d}\Phi/\mathrm{d}\tau)_{\tau=0}}{\Phi(0)
(\mu a)}
\notag
\\
& =\frac{10^3\kappa_m}{1.5\times 10^{-24}(\sqrt{2.85}\times 10^{15})}\simeq 40,
\end{align}
where we substitute the values of the parameters from Eqs.~(\ref{initial_Phi})-(\ref{mass2}) and put $\kappa_m=10^{-10}$.

In what follows the phase $\bar{\theta}= \varphi (t)/f_a=\Phi (\tau)/\Phi (0)$ oscillates and diminishes with the universe expansion
as $\sim t^{-3/4}$ (see Eq.~(45) in Ref.~\cite{Sikivie:2006ni}) for a zero mode evolution neglecting electromagnetism for axions, or as $\sim 1/a(\eta)\sqrt{a(\eta)}$ during radiation era, $a\sim \sqrt{t}$ 
(see 
Ref.~\cite[p.~40]{Gorbunov:2011zz}). Of course, the oscillation frequency $\omega(\tau)=\mu a(\tau)$ rises with the expansion that blackens oscillations in Fig.~\ref{fig:axion} for times $t\gg t_0$. Compare the insets in that plot where  oscillations with the large amplitude $\Phi/\Phi (0)\sim 40$ are distinguishable at $t\sim t_0$.

The more smooth evolution curves are seen in Fig.~\ref{fig:axionen} for the normalized axion energy density $\rho_a(t)=C_m\tilde{\rho}(\tau)$ given by Eqs.~\eqref{energy} and~\eqref{constant}. The ratio of $\rho_a(t)/\rho(t_0)$ corresponds to the ratio of dimensionless functions $\tilde{\rho}(\tau)$,
\begin{equation}\label{energy3}
\tilde{\rho}(\tau)=\left[\left(\frac{{\rm d}\Phi}{{\rm d}\tau}\right)^2 + \mu^2a^2\Phi^2(\tau)\right],
\end{equation}
and $\tilde{\rho}(0)$ plotted in Fig.~\ref{fig:axionen},
\begin{equation}\label{energy4}
\frac{\tilde{\rho}(\tau)}{\tilde{\rho}(0)}=\frac{({\rm d}\Phi/{\rm d}\tau)^2 + \mu^2a^2\Phi^2(\tau)}{({\rm d}\Phi/{\rm d}\tau)_{\tau=0}^2 + \mu^2a^2(0)\Phi^2(0)}.
\end{equation}
For $\kappa_m=10^{-10}$ the huge dimensional factor $C_m= (2/9)\times 10^{10}~{\rm GeV}^4$ in Eq.~\eqref{energy} is compensated in the product $\rho_a(t_0)=C_m\tilde{\rho}(\tau=0)$ by the small value $\tilde{\rho}(\tau=0)= 10^{-14}$ calculated numerically from the system of evolution Eqs.~(\ref{systemH})-(\ref{systemPhi}), or $\rho_a(t_0)=C_m\tilde{\rho}(\tau=0) =(2/9)\times 10^{-4}~{\rm GeV}^4$ is the initial axion energy density. On the other hand, the ultra-relativistic matter energy density $\rho=\pi^2T^4g^*/90$ in Friedmann's law  is given by the number of relativistic degrees of freedom $g^*=10.75$ and temperature $T_\mathrm{QCD}=0.1~{\rm GeV}$ at the same moment $t_0$, $\rho=1.18\times 10^{-4}~{\rm GeV}^4$. Thus, we get at the start a remarkable ratio of energy densities for the axion as a candidate to CDM, $\rho_a(t_0)/\rho=0.19$. Notice that this ratio does not depend on a width of the PMF Fourier spectrum $\kappa_m$ in Eq.~(\ref{width}) since that parameter shrinks in the initial axion energy density $\rho_a(t_0)=C_m\tilde{\rho}(0)$ where $C_m\sim \kappa_m^{-2}$ in Eq.~(\ref{constant}) and $\tilde{\rho}(0)\sim (d\Phi/d\tau)_{\tau=0}^2\sim \kappa_m^2$ in Eq.~(\ref{derivative_axion}).

\begin{figure*}
  \centering
  \subfigure[]
  {\label{fig:energy1}
  \includegraphics[scale=.33]{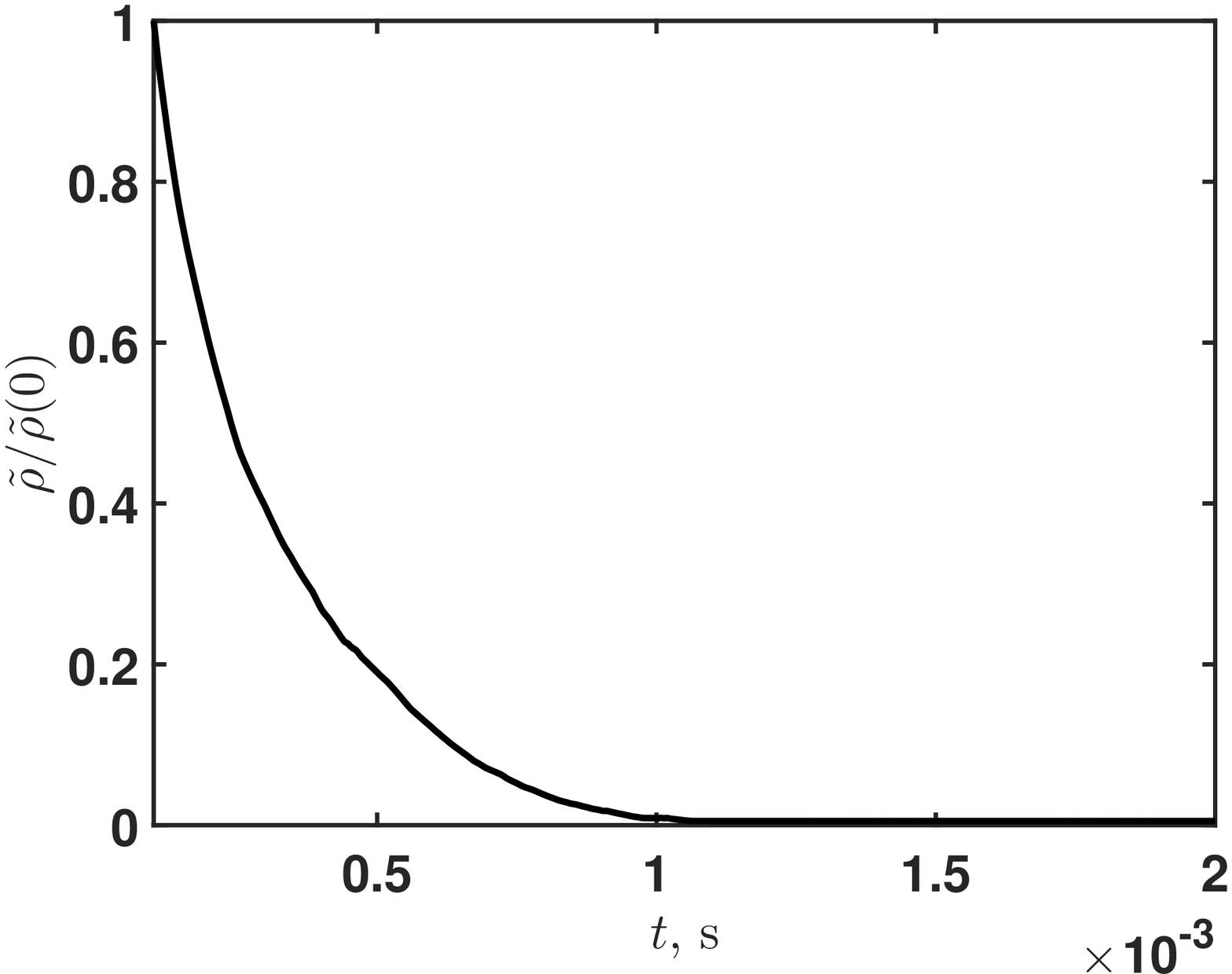}}
  \hskip-.6cm
  \subfigure[]
  {\label{fig:energy2}
  \includegraphics[scale=.33]{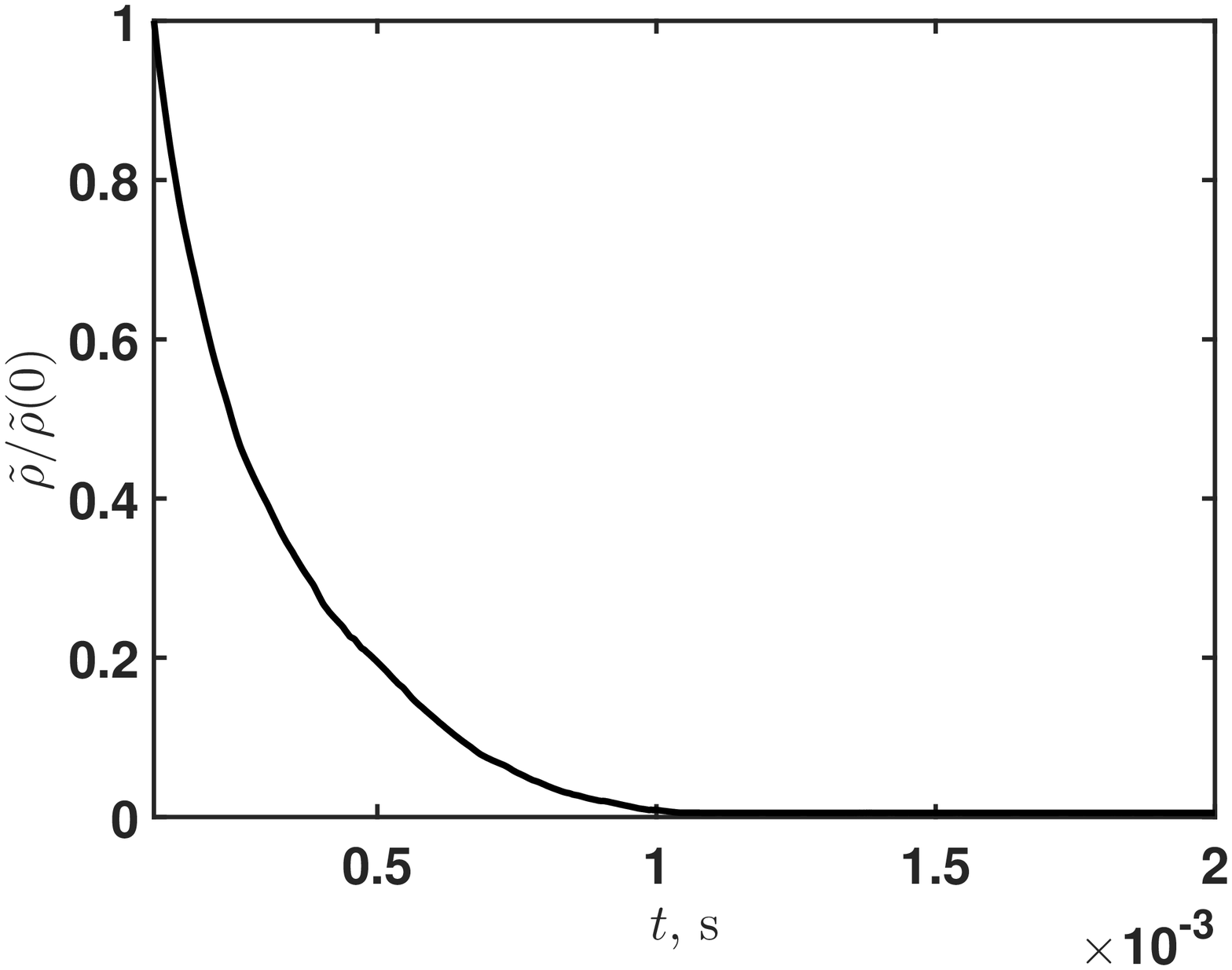}}
  \protect
  \caption{(a) The evolution of the axion energy density $\tilde{\rho}(\tau)/\tilde{\rho}(\tau=0)$ for the Peccei-Quinn parameter $f_\mathrm{PQ}\simeq f_a=10^{12}\,{\rm GeV}$ and the Fourier spectrum width  $\kappa_m= 10^{-10}$ in Eq.~(\ref{width}), choosing the maximal $B_\mathrm{now}=10^{-9}\,{\rm G}$. The seed magnetic field is supposed to be nonhelical, i.e. $q=0$ in Eq.~(\ref{initial_spectrum}). (b) The evolution of the axion energy amplitude $\tilde{\rho}(\tau)/\tilde{\rho}(\tau=0)$ for the Peccei-Quinn parameter $f_\mathrm{PQ}\simeq f_a=10^{12}\,{\rm GeV}$ without PMF contribution, $g_{a\gamma}=0$ and $q=0$.\label{fig:axionen}}
\end{figure*}

Then the axion energy density $\rho_a\sim \tilde{\rho}(\tau)$ reduces as seen in Fig.~\ref{fig:axionen} as well as the matter density $\rho$ decreases due to the universe cooling, $\rho\sim T^4$, therefore supporting approximately the same ratio $\rho_a(t)/\rho (t)\simeq 0.2$.

We are not guaranteed to observe in future the maximum PMF $B_\mathrm{now}=10^{-9}\,{\rm G}$ while some predictions with a lower bound for $B_{1\,{\rm Mpc}}(z=0) > 10^{-16}\,{\rm G}$  exist~\cite{Neronov:1900zz}. They arise, in particular, from $\gamma$-ray observation in the Fermi experiment~\cite{Fermi}. 

Recently, in Ref.~\cite{Jedamzik}, it was suggested to look for PMF at a level of the comoving strength $B_\mathrm{now}=Ba^2=\text{const}\simeq 10^{-11}\,{\rm G}$. In the last photon scattering (for which CMB instruments are sensitive), such a field corresponds to the PMF strength $B(z_r)=B_\mathrm{now}(1 + z_r)^2\simeq 10^{-5}\,{\rm G}$, whereas, in our problem, to $B_0=B_\mathrm{now}(1 +z_\mathrm{QCD})^2\simeq 1.6\times 10^{12}\,{\rm G}$.

\section{Discussion\label{sec:DISC}}

In the present work, we have studied the evolution of a zero axion mode, $\nabla\varphi=0$, in a hot plasma of the early Universe in the presence of PMF that do not exceed $B_\mathrm{now}= 10^{-9}\,{\rm G}$ at the present time. We conclude that such a mode can contribute to the axion CDM since its amplitude tends to $\bar{\theta}(t)=\varphi(t)/f_a\to 1$ evolving during radiation era, see Fig.~\ref{fig:axion}. Simultaneously the corresponding axion energy density $\rho_a(t)$ reduces in such a way that its part in the total matter energy density $\rho$ tends to the CDM ratio  $\rho_a(t)/\rho\simeq 0.2$, see comments on that after Eq. (\ref{energy4}) concerning the axion energy evolution in Fig.~\ref{fig:axionen}.

Notice that our PMF amplitudes $\sqrt{\int_{\kappa_m}^1 \mathrm{d}\kappa{\mathcal{R}}(\kappa,\tau)}\sim Ba^2=\text{const}$ correspond to the comoving strength for a given $B_\mathrm{now}=Ba^2=\text{const}$ and $\kappa_m$. Therefore, the ratio
\begin{equation}\label{PMF_constant}
\sqrt{\frac{\int_{\kappa_m}^1 \mathrm{d}\kappa {\mathcal{R}}(\kappa,\tau)}{\int_{\kappa_m}^1 \mathrm{d}\kappa {\mathcal{R}}(\kappa,\tau=0)}}\approx 1,
\end{equation}
does not depend on the conformal time $\eta\sim \tau$, being slightly violated by PMF interaction with axions.  The similar conservation concerns the magnetic helicity density, $\int_{\kappa_m}^1\tilde{\mathcal{H}}(\kappa,\tau)\mathrm{d}\kappa\sim ha^3=\text{const}$, calculated from the same self-consistent system of Eqs.~(\ref{systemH})-\eqref{systemPhi} for each given $B_\mathrm{now}$ and $\kappa_m$. Hence, the magnetic helicity density for the case $q=0$,
\begin{equation}\label{helicity_const}
\int_{\kappa_m}^1{\mathcal{H}}(\kappa,\tau)\mathrm{d}\kappa
\approx
\int_{\kappa_m}^1{\mathcal{H}}(\kappa,\tau=0)\mathrm{d}\kappa= 0,
\end{equation}
does not depend on time as well. The similar independence on time occurs in the case of a maximally helical PMF with $q=1$. For the PMF amplitude one obtains precisely  the result in Eq.~(\ref{PMF_constant}), while for the magnetic helicity density in Eq.~(\ref{helicity_const}) changes to
\begin{equation}\label{helicity_const2}
\int_{\kappa_m}^1 {\mathcal{H}}(\kappa,\tau)\mathrm{d}\kappa
\approx
\int_{\kappa_m}^1 {\mathcal{H}}(\kappa,\tau=0)\mathrm{d}\kappa = \text{const},
\end{equation}
where the constant depends on the initial conditions.

The conservation of the magnetic helicity density in Eq.~(\ref{helicity_const2}) is not surprising in the absence of the axion interaction with PMF's through Primakoff's effect. Indeed, the comoving mean energy density $\varepsilon_\mathrm{M}=\langle{\bf B}^2\rangle/2=\smallint \mathrm{d}k_c\rho_c(k_c,\eta)$ decays with time such that $\varepsilon_\mathrm{M}\propto\eta^{-2/3}$ while the correlation length grows like $\xi_\mathrm{M}\propto \eta^{2/3}$ as it follows from the dimensionless analysis based on the Kolmogorov-type approach for the magnetic turbulence, see, e.g., Refs.~\cite{Kahniashvili:2012uj,Brandenburg:2017neh}. This results in conservation of the magnetic helicity density, $h_c(\eta)=\smallint \mathrm{d}k_c h_c(k_c,\eta)=\langle{\bf B}^2\rangle\xi_\mathrm{M}=\text{const}$. If we switch additionally on the interaction of axions with PMF's, the violation of that conservation law is negligible because of a very small coupling constant $g_{a\gamma}=\alpha_\mathrm{em}/2\pi f_a$; cf. Eqs~(\ref{helicity_const}) and~(\ref{helicity_const2}).

Equations~\eqref{PMF_constant}-\eqref{helicity_const2} were directly checked by our numerical simulations. Thus, we confirm the conclusion of Ref.~\cite{Long:2015cza} that the axion zero mode ($\nabla\varphi=0$) practically  does not influence any PMF characteristics. The main discrepancies of our work and Ref.~\cite{Long:2015cza} consist in the exact accounting for the universe expansion in the dynamics of axions and electromagnetic fields. We have also taken into account the nontrivial spectra of the magnetic helicity and the magnetic energy densities.

On the first glance, there is a similarity of the axion electromagnetism with the chiral magnetic effect (CME) \cite{Boyarsky:2011uy,Tashiro:2012mf} that was a motivation for us to apply the Primakoff's mechanism and study the mutual influence of an axion zero mode and PMFs in the early Universe. 

However, it is not surprising that the CME is significantly more efficient for the amplification of PMFs than in the case of the axion interaction with magnetic fields. In the induction Eq.~(\ref{Faraday}), the helicity parameter $\alpha (t)=(g_{a\gamma}\mathrm{d}\varphi/\mathrm{d}t)/\sigma_\mathrm{cond}$, that enters there the PMF instability term $\alpha (t)(\nabla\times {\bf B})$, occurs to be much less than the corresponding CME parameter $\alpha_\mathrm{CME}(t)=2\alpha_\mathrm{em}\mu_5(t)/\pi\sigma_\mathrm{cond}$, where
$\mu_5=(\mu_\mathrm{R} - \mu_\mathrm{L})/2$ and $\mu_\mathrm{R,L}$ are the chemical potentials for right and left charged leptons. Indeed, the CME parameter, $\alpha_\mathrm{CME}$, is given by a great pseudoscalar $\mu_5/T\sim 10^{-5}$ at the temperature $T_\mathrm{QCD}\sim 10^8\,{\rm eV}$, or $\mu_5\sim 10^{3}\,{\rm eV}$ (see Fig.~1 in Ref.~\cite{Boyarsky:2011uy}). While in the case of axions the pseudoscalar $\alpha$, when accounting for the initial derivative $\mathrm{d}\varphi/\mathrm{d}t|_{t=t_0}\simeq m_a\varphi(t_0)=m_af_a$ [see below Eq.~(\ref{initial_Phi})], is estimated as 
\begin{equation}
\alpha=\left(\frac{\alpha_\mathrm{em}}{2\pi}\right)\frac{m_a}{\sigma_\mathrm{cond}}\ll \alpha_\mathrm{CME}=\left(\frac{2\alpha_\mathrm{em}}{\pi}\right)\frac{\mu_5}{\sigma_\mathrm{cond}},
\end{equation}
where, obviously, the axion mass  $m_a$ in Eq. (\ref{mass}) is quite small, $m_a\ll \mu_5$.

Thus,  we conclude that the axion zero mode, $\nabla\varphi=0$, and PMFs evolve independently each of other. Secondly, for reasonable PMF characteristics, the axion zero mode  that appears in the case when the inflation occurs after the Peccei-Quinn phase transition, $T_\mathrm{PQ}> T_\mathrm{R}$, $T_\mathrm{PQ}\simeq 10^{12}\,{\rm GeV}$, can be a candidate to the CDM. If there is no inflation after the PQ phase transition, $T_\mathrm{PQ}< T_\mathrm{R}$, the axion field is spatially varying, $\nabla \varphi\neq 0$. Axion strings can produce such nonzero momentum modes of the axion field which could contribute to the CDM~\cite{Sikivie:2006ni,Davidson:2014hfa}. The interaction of these modes with PMF is beyond scope of the present work.

\begin{acknowledgments}
We are grateful to V.~A.~Berezin for remarks concerning description of massive scalar fields in the general relativity with conformal variables.
\end{acknowledgments}

\appendix

\section{Complete set of evolution equations for axion zero mode and PMF spectra\label{A}}

In this appendix, we derive the equations for the spectra of the magnetic
energy and helicity, as well as for the axion field. 

The action for the axion field $\varphi$ with the mass $m_a$, interacting
with the electromagnetic field $A^{\mu}$, in curved spacetime is
\begin{align}
S= & \int\mathrm{d}^{4}x\sqrt{-g}\bigg[\frac{1}{2}\left(g^{\mu\nu}\partial_{\mu}\varphi\partial_{\nu}\varphi-m_a^{2}\varphi^{2}\right)-\frac{1}{4}F_{\mu\nu}F_{\lambda\rho}g^{\mu\lambda}g^{\nu\rho}\nonumber \\
 & -\frac{g_{a\gamma}\varphi}{4}F_{\mu\nu}\tilde{F}{}_{\lambda\rho}g^{\mu\lambda}g^{\nu\rho}-A_{\mu}J_{\nu}g^{\mu\nu}\bigg],\label{eq:action}
\end{align}
where $g=\det(g_{\mu\nu})$, $g_{\mu\nu}$ is the metric tensor, $F_{\mu\nu}=\partial_{\mu}A_{\nu}-\partial_{\nu}A_{\mu}$
is the electromagnetic stress tensor, $g_{a\gamma}$ is the coupling
constant, $\tilde{F}^{\mu\nu}=\tfrac{1}{2}E^{\mu\nu\alpha\beta}F_{\alpha\beta}$
is the dual counterpart of $F_{\mu\nu}$, $E^{\mu\nu\alpha\beta}=\tfrac{1}{\sqrt{-g}}\varepsilon^{\mu\nu\alpha\beta}$
is the covariant antisymmetric tensor, $\varepsilon^{0123}=+1$, and
$J^{\mu}$ is the external current.

Using Eq.~(\ref{eq:action}), one gets the equations for $\varphi$
and $A^{\mu}$ in the form, 
\begin{align}
\frac{1}{\sqrt{-g}}\partial_{\mu}\left(\sqrt{-g}\partial^{\mu}\varphi\right)+m_a^{2}\varphi+\frac{g_{a\gamma}}{4}F_{\mu\nu}\tilde{F}^{\mu\nu} & =0,\label{eq:eqphi}\\
\frac{1}{\sqrt{-g}}\partial_{\nu}(\sqrt{-g}F^{\mu\nu})+\partial_{\nu}\varphi\tilde{F}^{\mu\nu}+J^{\mu} & =0.\label{eq:eqA}
\end{align}
In Eq.~(\ref{eq:eqA}), we take into account that $\nabla_{\nu}\tilde{F}^{\mu\nu}=0$,
where $\nabla_{\mu}$ is the covariant derivative. This fact can be
proven basing on the Maxwell equation, $\nabla_{\mu}F_{\alpha\beta}+\nabla_{\alpha}F_{\beta\mu}+\nabla_{\beta}F_{\mu\alpha}=0$,
and the definition of $\tilde{F}^{\mu\nu}$.

Now we choose the FRW metric, which has the
form,
\begin{equation}
g_{\mu\nu}=\text{diag}(1,-a^{2},-a^{2},-a^{2}),
\end{equation}
where $a=a(t)$ is the scale factor. In this case, $\sqrt{-g}=a^{3}$.
We also take the zero mode approximation for the axion field, $\partial_{0}\varphi\neq0$
and $\partial_{i}\varphi=0$. The four-current $J^{\mu}$ in Eq.~(\ref{eq:eqA})
can be represented in the form, $J^{\mu}=(\rho,\mathbf{J}/a)$, where
$\rho$ is the charge density and $\mathbf{J}$ is the electric three-current.

Following Ref.~\cite{Brandenburg:1996fc}, we introduce the conformal variables,
\begin{equation}
\mathbf{E}_{c}=a^{2}\mathbf{E},\quad\mathbf{B}_{c}=a^{2}\mathbf{B},\quad\rho_{c}=a^{3}\rho,\quad\mathbf{J}_{c}=a^{3}\mathbf{J}.
\end{equation}
Then, Eq.~(\ref{eq:eqA}) takes the form,
\begin{align}
(\nabla\cdot\mathbf{E}_{c}) & =\rho_{c},\quad\frac{\partial\mathbf{E}_{c}}{\partial\eta}+\mathbf{J}_{c}+g_{a\gamma}\mathbf{B}_{c}\frac{\mathrm{d}\varphi}{\mathrm{d}\eta}=(\nabla\times\mathbf{B}_{c}),\nonumber \\
(\nabla\cdot\mathbf{B}_{c}) & =0,\quad\frac{\partial\mathbf{B}_{c}}{\partial\eta}=-(\nabla\times\mathbf{E}_{c}),\label{eq:Maxc}
\end{align}
where $\eta=\smallint\tfrac{\mathrm{d}t}{a}$ is the conformal time.
Equation~(\ref{eq:Maxc}) should be supplied with the relation between
$\mathbf{E}_{c}$ and $\mathbf{J}_{c}$: $\mathbf{J}_{c}=\sigma_{c}\mathbf{E}_{c}$,
where $\sigma_{c}=a\sigma=\text{const}$ and $\sigma$ is the conductivity.

After the volume integration, $V^{-1}\int \mathrm{d}^3x(...)$, Eq.~(\ref{eq:eqphi}) for the axion zero mode can be rewritten in the form,
\begin{equation}
\frac{\mathrm{d}^{2}\varphi}{\mathrm{d}\eta^{2}}+\frac{2}{a}\frac{\mathrm{d}a}{\mathrm{d}\eta}\frac{\mathrm{d}\varphi}{\mathrm{d}\eta}+m_a^{2}a^{2}\varphi=\frac{g_{a\gamma}}{Va^{2}}\int \mathrm{d}^3x (\mathbf{E}_{c}(\eta, {\bf x})\cdot\mathbf{B}_{c}(\eta,{\bf x})),\label{eq:phic}
\end{equation}
where we substituted $F_{\mu\nu}\tilde{F}^{\mu\nu}= -4({\bf E}{\bf B})$  multiplying Eq.~(\ref{eq:eqphi}) by $a^2$, and assumed that $\varphi$ does not change under the conformal transformations. The right hand side in Eq.~(\ref{eq:phic}) enters our master Eq.~(\ref{axion_wave}) as,
\begin{equation}\label{eq:hc}
-\frac{g_{a\gamma}}{2a^2}\frac{\mathrm{d}h_c(\eta)}{\mathrm{d}\eta}=\frac{g_{a\gamma}}{Va^{2}}\int \mathrm{d}^3x(\mathbf{E}_{c}(\eta, {\bf x})\cdot\mathbf{B}_{c}(\eta,{\bf x})).
\end{equation}
Such a change follows from the definition of the conformal PMF helicity density $h_c=a^3h=V^{-1}\int \mathrm{d}^3x{\bf A}_c(\eta,{\bf x})\cdot{\bf B}_c(\eta,{\bf x})$ obeying Maxwell Eq.~(\ref{eq:Maxc}). Here $h_c(\eta)=\int \mathrm{d}k_ch_c(k_c,\eta)$ is given by a spectrum of the PMF helicity.

We introduce the conformal spectra of the PMF energy and the PMF
helicity density,
\begin{align}
\rho_{c}(k_{c},\eta)=& \frac{k_{c}^{2}}{4\pi^{2}V}\mathbf{B}_{c}(k_{c},\eta)\cdot\mathbf{B}_{c}^{*}(k_{c},\eta),
\notag
\\
h_{c}(k_{c},\eta)=& \frac{k_{c}^{2}}{2\pi^{2}V}\mathbf{A}_{c}(k_{c},\eta)\cdot\mathbf{B}_{c}^{*}(k_{c},\eta),
\end{align}
where $k_{c}=ka$ is the conformal momentum, $\mathbf{A}_{c}(k_{c},\eta)$
and $\mathbf{B}_{c}(k_{c},\eta)$ are the Fourier components of the
vector potential and the magnetic field, and $V$ is the normalization
volume.

Basing on Eqs.~(\ref{eq:Maxc}) and~(\ref{eq:phic}), as well as
using the results of Ref.~\cite{Dvornikov:2013bca}, we get that,
\begin{align}
\frac{\partial h_{c}(k_{c},\eta)}{\partial\eta} & =-\frac{2k_{c}^{2}}{\sigma_{c}}h_{c}(k_{c},\eta)+\frac{4g_{a\gamma}}{\sigma_{c}}\frac{\mathrm{d}\varphi(\eta)}{\mathrm{d}\eta}\rho_{c}(k_{c},\eta),\nonumber \\
\frac{\partial\rho_{c}(k_{c},\eta)}{\partial\eta} & =-\frac{2k_{c}^{2}}{\sigma_{c}}\rho_{c}(k_{c},\eta)+k_{c}^{2}\frac{g_{a\gamma}}{\sigma_{c}}\frac{\mathrm{d}\varphi(\eta)}{\mathrm{d}\eta}h_{c}(k_{c},\eta),\label{eq:hrho}
\end{align}
and
\begin{align}
\frac{\mathrm{d}^{2}\varphi(\eta)}{\mathrm{d}\eta^{2}}+ & \frac{2}{a(\eta)}\frac{\mathrm{d}\varphi(\eta)}{\mathrm{d}\eta}\left(\frac{\mathrm{d}a(\eta)}{\mathrm{d}\eta}+\frac{g_{a\gamma}^{2}}{a(\eta)\sigma_{c}}\int_{k_{\mathrm{min}}}^{k_{\mathrm{max}}}\rho_{c}(k_{c},\eta)\mathrm{d}k_{c}\right)\nonumber \\
 & +m_a^{2}a^{2}(\eta)\varphi(\eta)=\frac{g_{a\gamma}}{a^{2}(\eta)\sigma_{c}}\int_{k_{\mathrm{min}}}^{k_{\mathrm{max}}}k_{c}^{2}h_{c}(k_{c},\eta)\mathrm{d}k_{c},\label{eq:phieta}
\end{align}
where $k_{\mathrm{min},\mathrm{max}}$ are the ranges of the $k_{c}$
variation.

Now we define the dimensionless variables,
\begin{equation}
\kappa=\frac{k_{c}}{k_{\mathrm{max}}},\quad\tau=\frac{2k_{\mathrm{max}}^{2}}{\sigma_{c}}\eta,
\end{equation}
and the dimensionless functions of these variables
\begin{align}
\mathcal{H}(\kappa, \tau)=& \left(\frac{g_{a\gamma}^2}{2}\right)
h_c(k_c,\eta),
\notag
\\
\mathcal{R}(\kappa,\tau)=&\left(\frac{g_{a\gamma}^2}{k_\mathrm{max}}\right)\rho_c(k_c,\eta),
\notag
\\
\Phi( \tau)=&\left(\frac{2k_\mathrm{max}g_{a\gamma}}{\sigma_c}\right)\varphi(\eta).
\end{align}
Equations~(\ref{eq:hrho}) and~(\ref{eq:phieta}) can be rewritten
in the form,
\begin{align}
\frac{\partial\mathcal{H}(\tau,\kappa)}{\partial\tau}= & -\kappa^{2}\mathcal{H}(\tau,\kappa)+\frac{\mathrm{d}\Phi(\tau)}{\mathrm{d}\tau}\mathcal{R}(\tau,\kappa),\nonumber \\
\frac{\partial\mathcal{R}(\tau,\kappa)}{\partial\tau}= & -\kappa^{2}\mathcal{R}(\tau,\kappa)+\kappa^{2}\frac{\mathrm{d}\Phi(\tau)}{\mathrm{d}\tau}\mathcal{H}(\tau,\kappa),\nonumber \\
\frac{\mathrm{d}^{2}\Phi(\tau)}{\mathrm{d}\tau^{2}}+ & \frac{2}{a(\tau)}\frac{\mathrm{d}\Phi(\tau)}{\mathrm{d}\tau}\left(\frac{\mathrm{d}a(\tau)}{\mathrm{d}\tau}+\frac{1}{2a(\tau)}\int_{\kappa_{m}}^{1}\mathcal{R}(\tau,\kappa)\mathrm{d}\kappa\right)\nonumber \\
 & =-\mu^{2}a^{2}(\tau)\Phi(\tau)+\frac{1}{a^{2}(\tau)}\int_{\kappa_{m}}^{1}\kappa^{2}\mathcal{H}(\tau,\kappa)\mathrm{d}\kappa,
\end{align}
where $\mu=m_a\sigma_{c}/2k_{\mathrm{max}}^{2}$ and $\kappa_{m}=k_{\mathrm{min}}/k_{\mathrm{max}}$.


\begin{thebibliography}{100}
\bibitem{f2}
Accounting for the scale factor $a=(1 + z)^{-1}$, we vary the  undetermined PMF strength $B_c=a^2B=B_\mathrm{now}$, where $B_\mathrm{now}(z=0)\leq 10^{-9}\,{\rm G}$ for extragalactic fields \cite{Plank}. For the initial time at the QCD phase transition the PMF occurs much stronger, $B_c(t_0)=B_\mathrm{now}(1 + z_\mathrm{QCD})^2$.
\bibitem{f4}
The equivalence to physical Fourier variables for the corresponding  PMF spectrum $\rho_\mathrm{B}(k,t_0)=C_1k^{n_\mathrm{B}}$ comes from its normalization on $B^2(t_0)/2=B_\mathrm{now}^2/2a^4(t_0)$ and the relation $\rho_c(\eta=0)=a^{4}(t_0)\rho_\mathrm{B}(t_0)$ for the conformal PMF energy density $\rho_c(\eta=0)$ and the standard one, $\rho_\mathrm{B}(t_0)=\int_{k_1}^{k_2}\mathrm{d}k\rho_\mathrm{B}(k,t_0)$, where $k$ are physical momenta, $k=k_c/a$.
\end{thebibliography}
\end{document}